\documentclass[twocolumn]{aastex62}
\usepackage{graphics,epsf}
\usepackage{amsmath}                
\usepackage{amsfonts}               
\usepackage{amssymb}                
\usepackage{epsfig}                 
\usepackage{graphicx}
\usepackage{float}
\usepackage{color}
\usepackage[para,online,flushleft]{threeparttable}

\newcommand{\cm}{{~\rm cm}}
\newcommand{\km}{{~\rm km}}
\newcommand{\s}{{~\rm s}}

\newcommand{\g}{{~\rm g}}

\newcommand{\erg}{{~\rm erg}}

\newcommand{\days}{{~\rm days}}

\begin{document}

\title{Jet-shaped geometrically modified light curves of core collapse supernovae} 


\author{Noa Kaplan}
\affiliation{Department of Physics, Technion, Haifa, 3200003, Israel; noa1kaplan@campus.technion.ac.il; soker@physics.technion.ac.il}

\author[0000-0003-0375-8987]{Noam Soker}
\affiliation{Department of Physics, Technion, Haifa, 3200003, Israel; noa1kaplan@campus.technion.ac.il; soker@physics.technion.ac.il}
\affiliation{Guangdong Technion Israel Institute of Technology, Shantou 515069, Guangdong Province, China}

\begin{abstract}
We build three simple bipolar ejecta models for core collapse supernovae (CCSNe), as expected when the explosion is driven by strong jets, and show that for an observer located in the equatorial plane of the ejecta, the light curve has a rapid luminosity decline, and even an abrupt drop. 
In calculating the geometrically modified photosphere we assume that the ejecta has an axisymmetrical structure composed of an equatorial ejecta and faster polar ejecta, and has a uniform effective temperature. At early times the photosphere in the polar ejecta grows faster than the equatorial one, leading to higher luminosity relative to a spherical explosion. The origin of the extra radiated energy is the jets. At later times the optical depth decreases faster in the polar ejecta, and the polar photosphere becomes hidden behind the equatorial ejecta for an observer in the equatorial plane, leading to a rapid luminosity decline. For a model where the jets inflate two low-density polar bubbles, the luminosity decline might be abrupt.  
This model enables us to fit the abrupt decline in the light curve of SN~2018don.
\end{abstract}


\section{Introduction}
\label{sec:intro}

The collapsing core in core-collapse supernovae (CCSNe) liberates a huge amount of gravitational energy as it forms a neutron star (NS) remnant. If outer parts of the core, and possibly also parts of the envelope, collapse as well, then the remnant is a black hole (BH). Neutrinos from the hot newly born NS carry most of this energy within seconds. An explosion mechanism utilizes only a small fraction of the gravitational energy to explode the rest of the core and the envelope (or only the rest of the envelope in case the remnant is a BH).  
There is no agreement yet on the exact explosion mechanism.  
   
One explosion mechanism is driven by neutrinos that heat the in-flowing gas, i.e., the \textit{delayed neutrino mechanism} \citep{BetheWilson1985}. The other explosion mechanism is driven by jets that the newly born NS or BH launch, even in the majority of cases where the total angular momentum of the collapsing core is small, i.e., far too little to form a sustained accretion disk around the central remnant. The key process is the accretion of material with stochastic angular momentum, such that it forms a stochastic intermittent accretion torus (or disk) that launches intermittent jets with varying directions. This is the \textit{jittering jets explosion mechanism} (e.g., \citealt{Soker2010, PapishSoker2011, GilkisSoker2014, Quataertetal2019}). 
 
These two mechanisms have some difficulties that require additional ingredients for their solution. Simulations of the delayed neutrino mechanism regularly contain now convection or other perturbations in the pre-collapse core (e.g., \citealt{CouchOtt2013, Mulleretal2019Jittering}). This additional ingredient of pre-collapse core convection leads to relatively large variations in the magnitude and direction of the angular momentum of the mass that the newly born NS accretes. 
As a result of that, a number of three-dimensional core collapse simulations lead to outflow structures that resemble jittering jets, e.g., the bipolar outflow changes its direction \citep{Soker2019JitSim}. Namely, the extra ingredient that might solve some of the problems of the delayed neutrino mechanism (but not all problems, e.g., \citealt{SawadaMaeda2019}) leads to jittering jets. 
In the jittering jets explosion mechanism the extra ingredient that might help solve some problems is neutrino heating \citep{Soker2018KeyRoleB, Soker2019SASI}. 
      
The jittering jets explosion mechanism is applicable to all CCSNe. It predicts that even when the remnant is a BH, the final envelope material to be accreted launches jittering jets that explode the star (e.g., \citealt{GilkisSoker2014, GilkisSoker2015, Quataertetal2019}). Namely, there are no failed CCSNe according to the jittering jets explosion mechanism. To the contrary, the formation of a black hole might lead to super-energetic CCSNe with explosion energies up to $E_{\rm exp} > 10^{52} \erg$ \citep{Gilkisetal2016Super}. There are other different predictions of the two explosion mechanisms (e.g., \citealt{Gofmanetal2020}). 
   
{{{{ The jittering jets explosion mechanism faces several challenges. Most challenging to overcome is the requirement of the formation of intermittent accretion disks or belts. Three-dimensional simulations do not form such disk/belts (e.g., \citealt{Mulleretal2017}). The inclusion of magnetic fields and proper pre-collapse convection might help the formation of intermittent disks/belts that launch jets \citep{Soker2018KeyRoleB}. For the present study, another challenge is the expansion of the two opposite jets to large distance. \cite{Mostaetal2014} find that jets with strong magnetic fields suffer from the kink instability that prevents their expansion to large distances. This is not a general problem for the jittering jets explosion mechanism where the jets are strongly shocked at a distance of only ${\rm few} \times 1000 \km$ from the center and inflate large high-pressure bubbles (lobes) that explode the stars (e.g., \citealt{PapishSoker2014}). But the disruption of the jets might be a challenge to the present study, where we require the jets to expand to large distances and inflate two large polar bubbles in the ejecta. }}}}
  
In its prediction that jets might explode all CCSNe, the jittering jets explosion mechanism is significantly different from most other jet-driven mechanisms that require the pre-collapse core to be rapidly rotating, hence these are applicable only to rare cases (e.g., \citealt{Khokhlovetal1999, Aloyetal2000, Hoflich2001, Burrows2007, Nagakuraetal2011, TakiwakiKotake2011, Lazzati2012, Maedaetal2012, Bromberg_jet, Mostaetal2014, LopezCamaraetal2014, BrombergTchekhovskoy2016, LopezCamaraetal2016, Nishimuraetal2017, Fengetal2018, Gilkis2018}). 
 Although the explosion occurs in a time of a few seconds or less, late fall back onto the compact remnant might lead to very late, up to months after explosion, jets (e.g., \citealt{KaplanSoker2020}). 
 
The morphologies of many core collapse supernovae (CCSN), such as two opposite protrusions (`Ears'), and the observed polarization in some CCSNe, can be accounted for by jets (e.g., \citealt{Wangetal2001, Maundetal2007, Milisavljevic2013, Gonzalezetal2014, Marguttietal2014, Inserraetal2016, Mauerhanetal2017, GrichenerSoker2017, Bearetal2017, Garciaetal2017,  LopezFesen2018}). We take the view that jets play a major role in these, and possibly in most, CCSNe. 
There are other indications supporting bipolar explosion morphologies. 
For example, \cite{Boseetal2019} argue, based on the nebular-phase Balmer emission, that the $^{56}$Ni in the Type II-P CCSN SN~2016X (ASASSN-16at) has a bipolar morphology. Jet-driven CCSN explosions can form bipolar morphological features (e.g., \citealt{Orlandoetal2016, BearSoker2018}).
 
In the present study we examine one possible implication of bipolar explosions. Namely, an explosion where the ejecta has two opposite bubbles (lobes) along the symmetry axis. This is a rare case where the jets of the last jets-launching episode are very strong. In cases of observed `Ears' in CCSN remnants, the energy of the jets that inflate the Ears is only $\approx 1 - 10 \%$ of that of the CCSN \citep{GrichenerSoker2017}. We consider much stronger jets that inflate lobes/bubbles (i.e., very large Ears).  Non-spherical circumstellar matter (CSM) can affect the properties of the CCSN emission (e.g., \citealt{Soumagnacetal2019, Soumagnacetal2020}). We consider non-spherical ejecta, and do not refer to a CSM. In section \ref{sec:model} we describe our very simple model of the ejecta. In section \ref{sec:light_curve} we describe the effects of the highly non-spherical ejecta on the light curve that an observer in the equatorial plane measure. In section \ref{sec:Abrupt_drop} we build a more extreme model and try to explain the abrupt drop in the light curve of SN~2018don. 
We summarize our results and conclusions in section \ref{sec:summary}.

\section{Constructing the bipolar ejecta}
\label{sec:model}
\subsection{Basic properties}
\label{subsec:general_model}

We present an axisymmetrical SN explosion model with faster polar velocities, and try to explain a rapid drop in the light curve due to this geometry for an observer in the equatorial plane. Jets that the central newly born NS (or BH) launches might lead to such a geometry, as well as specific kinds of instabilities in the explosions.
Here we take jets to inflate the two opposite polar lobes. 
We assume that jets add energy and remove mass from the polar directions (to the sides), such that the polar regions expand faster, and the gas there might be hotter. 
Initially the photosphere moves faster in the polar directions. Later, as the material in the outskirts of the ejecta becomes optically thin, the result of the faster outflow and lower mass in the polar directions is a much lower density, and therefore the photosphere there retreats faster than in equatorial ejecta. Eventually, the photosphere in the equatorial ejecta is at larger radius than that in the polar ejecta.  

We build a very simple geometrical model that gives the shape of the photosphere. We assume an axisymmetrical structure where the polar outflow and the equatorial outflow each expands like part of a spherical explosion, but with different mass and energy. The polar ejecta has a half-opening angle (measured from the symmetry axis) of $\theta_0$. It expands as if it is part of a spherical explosion of mass $M_{\rm po}$ and an energy of $E_{\rm po}$. The actual mass and energy of the polar ejecta (together in both sides of the equatorial plane) are 
\begin{equation}
M_{\rm po,\theta_0}=(1-\cos \theta_0) M_{\rm po}; \quad  
E_{\rm po,\theta_0}=(1-\cos \theta_0) E_{\rm po},
\label{eq:MEpotheta0}
\end{equation}
 respectively. 
The equatorial ejecta expands in a section with an angle from the equatorial plane of $-(90^\circ - \theta_0)$ to $90^\circ - \theta_0$, as if it is a part of a spherical explosion with a mass of $M_{\rm eq}$ and an energy of $E_{\rm eq}$. The actual mass and energy of the equatorial ejecta are 
\begin{equation}
M_{\rm eq,\theta_0}=\cos \theta_0 M_{\rm eq}; \quad  
E_{\rm eq,\theta_0}=\cos \theta_0 E_{\rm eq},
\label{eq:MEeqtheta0}
\end{equation}
 respectively. 

We take the density profile of a spherical explosion of energy $E$ and mass $M$ from \cite{Chevalier_Soker89}
\begin{equation}
\rho = \begin{cases}
        K t^{-3} E^{-3/2} M^{5/2} \left( \frac{r}{t v_{\rm br}} \right)^{-1} 
        & r\leq t v_{\rm br}
        \\
        K t^{-3} E^{-3/2} M^{5/2} \left( \frac{r}{t v_{\rm br}} \right)^{-10} 
        & r>t v_{\rm br}.
        \end{cases}
\label{eq:density_profile}
\end{equation}
The velocity where the power law changes its value (i.e., the break in the density profile) is $v_{\rm br}=1.69({E}/{M})^{1/2}$.
We then calculate the location of the photosphere along each radial direction, $r_i$, from 
\begin{equation}
\tau = \int_{r_i}^\infty \kappa \rho dr = \frac{2}{3}.
\label{eq:tau}
\end{equation}
Since the density profile is a broken power law that changes its slope at a velocity of $v_{\rm br}$, we first perform the integration for the outer part of the power law, i.e., for $r>v_{\rm br} t$. If it is optically thick, then the photosphere is at $r_i=r_1>v_{\rm br} t$, where  
\begin{eqnarray}
\begin{aligned} 
& r_1 = 2.6 \times 10^{10} 
    \left(\frac{\kappa}{0.03 \cm^2 \g^{-1}} \right)^{1/9} 
    \left(\frac{E}{4 \times 10^{51} \erg}\right)^{7/18}
   \\&  \times 
    \left(\frac{M}{6 M_{\odot}}\right)^{-5/18}
    \left(\frac{t}{1 \s}\right)^{7/9} \cm;  \qquad {\rm for} \quad \tau (v_{\rm br} t) > 2/3 . 
\end{aligned}
\label{eq:r1}
\end{eqnarray}
If the outer part of the ejecta is optically thin, i.e., $\tau (v_{\rm br} t) < 2/3$, we neglect the contribution of the outer part (gas at $r>v_{\rm br} t$), and consider only the contribution of the inner part of the power law to the optical depth. This gives the photosphere at $r_i =r_2$, where  
\begin{eqnarray}
\begin{aligned} 
r_2 =  & 9.8  \times 10^8  
    \left(\frac{E}{4 \times 10^{51} \erg}\right)^{1/2}  
    \left(\frac{M}{6 M_{\odot}}\right)^{-1/2} \left(\frac{t}{1 \s}\right) 
    \\ \times & \exp 
    \left[ - 1.4 \times 10^{-14} 
    \left(\frac{\kappa}{0.03 \cm^2 \g^{-1}}\right)^{-1} \right.
    \\ & \times \left.
    \left(\frac{E}{4 \times 10^{51} \erg}\right) 
    \left(\frac{M}{6 M_{\odot}}\right)^{-2} \left(\frac{t}{1 \s}\right)^2 \right] \cm.
\end{aligned}
\label{eq:r2}
\end{eqnarray}

\subsection{Evolution of the photosphere}
\label{subsec:photosphere}

We now present the shape of the photosphere for one  axisymmetrical explosion geometry. We take the following parameters for the two spherical explosions that determine the properties of the equatorial and polar ejecta (recall that the actual masses and energies are according to equations \ref{eq:MEpotheta0} and \ref{eq:MEeqtheta0})  
\begin{eqnarray}
\begin{aligned} 
& E_{\rm eq} = 4 \times 10^{51} \erg; \qquad & M_{\rm eq} = 6 M_\odot, \\
& E_{\rm po} = \eta_{\rm E} E_{\rm eq};  \qquad 
& M_{\rm po} = \eta_{\rm M} M_{\rm eq}, 
\end{aligned}
\label{eq:E_M_th}
\end{eqnarray}
where the last line defines $\eta_{\rm E}$ and $\eta_{\rm M}$. Here we take $\eta_{\rm E}=2$ and $\eta_{\rm M}=0.5$. 
{{{{We also take in this case $\theta_0 = 60^\circ$ such that $(1-\cos\theta_0)=1/2$, which implies (using equations \ref{eq:MEpotheta0} and \ref{eq:MEeqtheta0}) that the actual total energy and mass of the CCSN are $E_{\rm SN} = (1-\cos{\theta_0})E_{\rm po}+\cos{\theta_0}\;E_{\rm eq} = 6 \times 10^{51} \erg$, and $M_{\rm SN}= (1-\cos{\theta_0})M_{\rm po}+\cos{\theta_0}\;M_{\rm eq} = 4.5 M_\odot$, respectively.}}}} We will later refer to the value of $E_{\rm eq}$ in equation (\ref{eq:E_M_th}) as the high-energy case, {{{{and below we will also deal with the low-energy case having $E_{\rm eq} = 0.667 \times 10^{51} \erg$.}}}}
 
\begin{figure}[ht!]
	\centering
	\includegraphics[trim=5.4cm 8.2cm 4cm 8.2cm ,clip, scale=0.8]{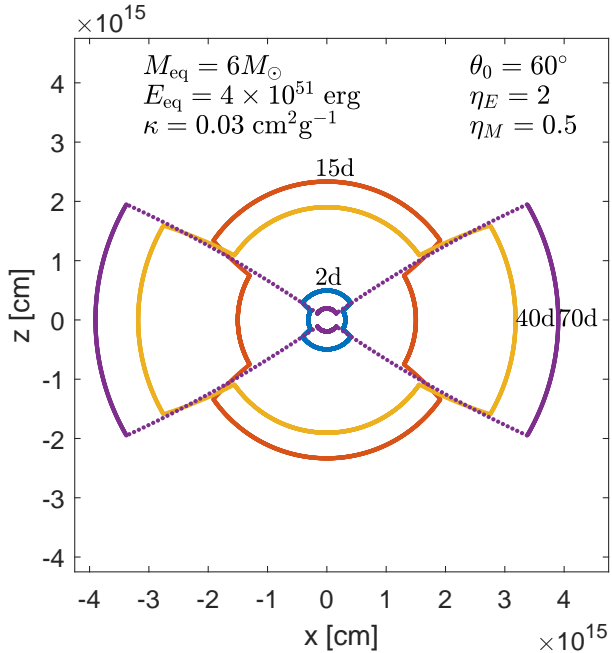}
	\caption{
	The shape of the photosphere in the meridional plane (a plane through the center and perpendicular to the equatorial plane; here the $y=0$ plane) for the high-energy case (equation \ref{eq:E_M_th}). The different colors depict the photosphere at different times, as labeled in days. At early times (here $2$ and $15 \days$) the photosphere in the polar ejecta (along and near the symmetry axis $x=y=0$) expands faster than the photosphere in the equatorial ejecta. At later times (here $40$ and $70 \days$) the outer parts of the polar ejecta becomes optically thin, and the photosphere in the polar directions rapidly recedes.
	}
	\label{fig:shapes_early_times}
\end{figure}

In Fig.~\ref{fig:shapes_early_times} we present the location of the photosphere in the meridional plane $y=0$ as we calculate from equations (\ref{eq:r1}) and (\ref{eq:r2}) at four times. At early times the photosphere is larger in the polar directions (the gas within $\theta_0 =60^\circ$ from the symmetry axis $x=y=0$) because the polar ejecta moves faster. As the optical depth decreases, the photosphere moves inward in the polar ejecta. As a result of that the polar ejecta becomes optically thin and cools earlier than the equatorial ejecta. At late times ($t=40 \days$ and $t=70 \days$ in the figure) the polar ejecta is already optically thin. 
At early times both the polar and equatorial radius of the photosphere is $r_1$. At about $10 \days$ 
the polar radius becomes $r_2$, and at later times (about $30 \days$) 
 the equatorial radius also becomes $r_2$.

\section{The light curve}
\label{sec:light_curve}

\subsection{Methods}
\label{subsec:MEthods}
  
{{{{ Before we present our results, we briefly summarize our method for calculating the light curve.  
We first build (section \ref{sec:model}, where we give more details) a simple (toy) geometrical model from two components, the equatorial ejecta and the polar ejecta (Fig. \ref{fig:shapes_early_times}).   
The polar ejecta is within an angle $\theta < \theta_0$ from the polar (symmetry) axis; we take in this study $\theta_0=60^\circ$ for two models (three cases), and $\theta_0=45^\circ$ for the third model. This structure is based in part on the morphologies of several bipolar planetary nebulae that are likely shaped by jets. These show polar lobes with lower density and larger radii (distance form the center), with a relatively sharp boundary between the two components, as we assume here at an angle $\theta_0$. A good example is the
Dumbbell planetary nebula (NGC~6853, M27, PN~G060.8-03.6; for images, e.g., \citealt{Meabornetal2005}\footnote{For a good image see also, e.g., 
https://www.nasa.gov/feature/goddard/2017/messier-27-the-dumbbell-nebula.})
. }}}}

{{{{ We then assume that the entire photosphere, of both components, has a uniform effective temperature (that varies with time). This implies that the relevant luminosity of the ejecta for an observer is the cross section (projection) of the ejecta on the plane of the sky and the effective temperature. For two SNe having the same effective temperature at a given time, the ratio of luminosities equals the ratio of the cross sections of their respective ejecta on the plane of the sky. We elaborate on this in section \ref{subsec:geometriccal} below for an equatorial-plane  observer. }}}}
 
{{{{ We do not calculate the light curve of the SN models we use from zero. We rather use a template SN light curve (section \ref{subsec:lightcurve}). We then use the ratio of the cross section of the ejecta of each of our four cases (of three geometrical models; sections \ref{subsec:lightcurve}, \ref{subsec:Shorter_fall}, and  \ref{sec:Abrupt_drop}, respectively) to that of a spherical ejecta, and together with the template SN light curve we calculate the light curves of the geometrically modified light curves. For the calculation of the cross section ratios  (projected area ratios) and then the light curves, we use MATLAB. }}}}
   
\subsection{The geometrical factor}
\label{subsec:geometriccal}

We consider an observer in the equatorial plane, and a black body emission by the ejecta with a uniform effective temperature on the photosphere. In such a case the effective luminosity is
\begin{eqnarray}
\begin{aligned} 
L_{\rm eff} (t) = 4 A_{\rm cross}(t) T^4_{\rm eff}(t),
\end{aligned}
\label{eq:Leff}
\end{eqnarray}
where $A_{\rm cross}$ is the cross section of the ejecta photosphere as seen by the observer. 
By cross section we refer to the projection of the photosphere on the plane of the sky. 
Namely, the flux arriving on Earth is $\phi=L_{\rm eff} / (4 \pi D^2)$, where $D$ is the distance to the CCSN. For example, at $t=15 \days$ for the case presented in Fig.~\ref{fig:shapes_early_times}, the value of $A_{\rm cross}(15 \days)$ is the area enclosed by the red line on the plane of the figure. 
At $t=70 \days$ the relevant cross section for an observer in the equatorial plane is a band on the plane of the sky, delineated on the sides by the two purple arcs, and by two horizontal lines, one connecting the upper edges of the two arcs {{{{and one connecting the two lower edges of the purple arcs.}}}} Namely, after the rapid collapse of the polar photosphere, an observer located in the equatorial plane will see a shape that resembles the side view of a torus. In this case and for $\theta_0=60^\circ$ it has a cross section with an area that is $0.61$ times that of a sphere with the same radius. Most of the radiation from the flattened ejecta is emitted now in the polar directions. This causes a drop in the light curve for an equatorial-plane observer.
    
\begin{figure}[ht!]
	\centering
	\includegraphics[trim=3.5cm 8cm 4cm 8cm ,clip, scale=0.6]{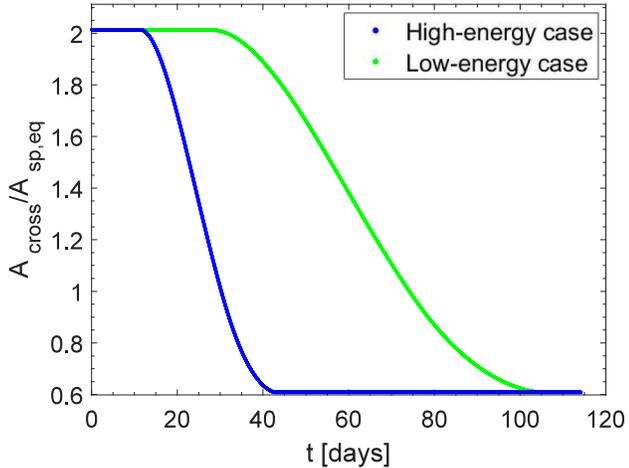}
	\caption{The ratio of the cross sections of the geometrically-modified SN photosphere to the cross section of a spherical explosion with radius equal to that of the equatorial ejecta, $A_{\rm cross}/A_{\rm sp,eq}$. By cross section we refer here to the projection of the photosphere on the plane of the sky as an observer in the equatorial plane ($z=0$) sees. 
	The blue (left) line represents the high-energy case with the same parameters as in Fig.~\ref{fig:shapes_early_times} (see inset there). The green (right) line represents the low-energy case with energies of $E_{\rm eq} = 0.667 \times 10^{51} \erg$ and $E_{\rm pol} = 1.33 \times 10^{51} \erg$; all the remaining parameters are as in the high-energy case. 
	}
	\label{fig:area_ratio}
\end{figure}

Overall, Fig.~\ref{fig:shapes_early_times} shows that the polar outflow increases the area at early times relative to that of a sphere with a radius as that of the equatorial ejecta, $A_{\rm sp,eq}\equiv \pi R^2_{\rm eq}$, where $R_{\rm eq}$ is the photospheric radius of the equatorial ejecta. Namely, at early times $A_{\rm cross} > A_{\rm sp,eq}$. Later the polar outflow becomes optically thin and the cross section decreases, $A_{\rm cross} < A_{\rm sp,eq}$. To quantify this ratio, in Fig.~\ref{fig:area_ratio} we present the evolution of the ratio $A_{\rm cross}(t)/A_{\rm sp,eq}(t)$ with time, for our high-energy and low-energy cases.
The high-energy case has the same parameters as the photosphere we present in Fig.~\ref{fig:shapes_early_times}, with $E_{\rm eq} = 4 \times 10^{51} \erg$. In the low-energy case we change the equatorial energy to be $E_{\rm eq} = 0.667 \times 10^{51} \erg$. Because we keep $\eta_E=2$, the polar energy decreases by the same ratio relative to the high-energy case, to be $E_{\rm pol} = 1.33 \times 10^{51} \erg$. All other parameters are as in the high-energy case.

\subsection{geometrically-modified light curves}
\label{subsec:lightcurve}

\begin{figure}[ht!]
	\centering
\includegraphics[trim=3.5cm 8cm 4cm 8cm ,clip, scale=0.6]{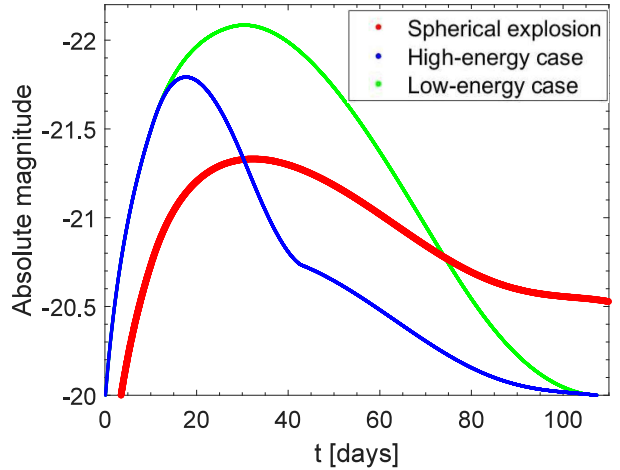}
\caption{Light curves for an observer in the equatorial plane (the $z=0$ plane in Fig.~\ref{fig:shapes_early_times}). 
The thick-red line is the light curve of a spherical explosion having the parameters we use for the equatorial ejecta of each of the two cases (high-energy and low-energy), $L_{\rm sp,eq}$. We assume that the light curve of the spherical explosion of the two cases is the same.
The blue (left) and green (right) lines are the geometrically-modified light curves {{{{of the high-energy and low-energy cases, respectively,}}}} as we calculate from equation (\ref{eq:lum}).
}
\label{fig:abs_magnitude}
\end{figure}

We start with a template light curve. We chose, somewhat arbitrary (we could have taken other light curves), the light curve of SN~2007bi that we take from The Supernova Catalog \cite{SN_catalog}.
We take this light curve to be that of a spherical explosion with the parameters we use here for the equatorial ejecta, i.e., we take $L_{\rm sp,eq}=L({\rm SN~2007bi})$. We present this light curve (in magnitude) by the thick-red line in Fig.~\ref{fig:abs_magnitude}. 

The luminosity of the geometrically-modified SNe ejecta,  under the assumption of a uniform photospheric temperature, is given by 
\begin{eqnarray}
L_{\rm SN}(t) = L_{\rm sp,eq} \frac{A_{\rm cross}}{A_{\rm sp,eq}}. 
\label{eq:lum}
\end{eqnarray}
  
We calculate two cases for this geometrical model. In the high-energy case, that we present the photosphere of in Fig.~\ref{fig:shapes_early_times}, we take $E_{\rm eq} = 4 \times 10^{51} \erg$ and $E_{\rm pol} = 8 \times 10^{51} \erg$. In the low-energy case we take $E_{\rm eq} = 0.667 \times 10^{51} \erg$ and $E_{\rm pol} = 1.33 \times 10^{51} \erg$, keeping all other parameters as in the high-energy case.  
We assume that for both cases the spherical light curve, $L_{\rm sp,eq}(t)$, is the same (thick-red line in Fig.~\ref{fig:abs_magnitude}). That we take the same light curve for the two cases that differ by a factor of six in their energies, means that the radiation carries the same amount of energy in the two cases for a spherical explosion (had all gas expanded like the equatorial ejecta). Although the low-energy case is six times less energetic, its slower expansion relative to the high-energy case implies that adiabatic loses are slower, and therefore photons have more time to diffuse out and carry a larger fraction of the energy. Taking the same $L_{\rm sp,eq}(t)$ for both cases allows a clear comparison between them.
   
We present the light curve of the high-energy case by a blue line (that has a rapid fall at about 20 days) in Fig.~\ref{fig:abs_magnitude}, while we present the light curve of the low-energy case with a green line, that falls more gradually and at later times. 
 
The geometrically-modified SN light curves that we study in this section have the following properties in relation to the non-modified SN light curve (thick-red lines in Fig.~\ref{fig:abs_magnitude}).
(1) A more rapid rise to maximum. (2) Higher luminosity at maximum. 
These two properties are related to our assumption of a uniform photospheric (effective) temperature. 
If there is no extra energy source, the faster expanding polar ejecta will be cooler than the equatorial ejecta. This is for two reasons, the faster adiabatic cooling of the faster ejecta and the more rapid photon diffusion (because of lower density). However, in our scenario there are two opposite jets that the newly born NS (or BH) launches along the polar directions that shape the bipolar outflow. These jets are expected to supply kinetic and thermal energy to the polar outflow, such that it might actually be hotter than the equatorial outflow \citep{KaplanSoker2020}.  

(3) The third property of the geometrically-modified light curves is a more rapid decline from maximum. This rapid decline occurs as the cross section of the photosphere of the polar ejecta decreases and finally disappears from the view of an observer near the equatorial plane. If indeed the photosphere of the polar ejecta is hotter than that of the equatorial ejecta (because of the jet interaction), then the decline in the light curve after the peek will be more rapid even. Most of the radiation from the polar ejecta after decline is emitted toward the polar directions and therefore do not reach an observer near the equatorial plane.

(4) For higher energies the effect of the polar lobes occurs earlier, and so the rapid drop occurs earlier. {{{{Specifically, for a polar ejecta with a higher energy the radius of the photosphere (r1 and r2) increases faster, hence the density is lower and so optical depth is lower and the radius of the photosphere decreases faster. In this case the radius of the photosphere changes from r1 to r2 earlier (earlier effect of the polar lobes) and r2 drops faster (drop occurs earlier).}}}} If this rapid drop occurs before the decline of the spherical explosion the effect is more pronounced. This is evident from the break at $t \simeq 40 \days$ in the high-energy case light curve (blue line), where after the rapid luminosity decline the decline returns to be more gradual. 
   
\subsection{A late rapid luminosity fall}
\label{subsec:Shorter_fall}

We construct a bipolar ejecta model where the jets shape the ejecta weeks after the explosion. Such jets result from a late fall-back onto the newly born NS (or BH). We considered such a late jet-ejecta interaction that inflate bipolar bubbles in \cite{KaplanSoker2020}. Here we consider a stronger jet-ejecta interaction where the jet-inflated bubbles break out from the ejecta. 
As we show now, this leads to a rapid luminosity fall at a later time as compared to early-shaped bipolar ejecta that we study in Sections \ref{subsec:geometriccal} and \ref{subsec:lightcurve}. We refer to this case as the late-fall case. 
   
In this case the ejecta maintains a spherical homologous expansion at early times. Here we take the explosion energy and mass for this spherical-expansion phase to be $E_{\rm eq} = 0.667 \times 10^{51} \erg$ and $M_{\rm eq} = 8 M_\odot$, respectively. 
We assume that the compact remnant at the center accretes mass at about one month after the explosion and launches two opposite jets. We assume also that these fast jets catch up with the ejecta and inflate two bubbles that shape the ejecta to have a bipolar morphology; the bipolar morphology starts to evolve about ten days later. 

We take the following set of parameters to construct the bipolar morphology of the late-fall case. We assume that at $t_{\rm bipol} = 40 \days$ the two opposite polar bubbles start to expand with a half opening angle of $\theta_0=60^{\circ}$, as measured from the symmetry (polar) axis. They expand in the same way as a spherical explosion with an energy of $E_{\rm po} = 1.33 \times 10^{51} \erg$ and $M_{\rm po} = 2.4 M_\odot$ would expand (but only in the cone of $\theta_0=60^{\circ}$ from the symmetry axis; see Section \ref{subsec:geometriccal}). Namely, we take $\eta_{\rm E}= 2 $ and $\eta_{\rm M} = 0.3$ in equation \ref{eq:E_M_th}. 
To have the polar ejecta radius to be the same as the equatorial ejecta at $t=40 \days$, we set the explosion that mimics the polar ejecta to occur at $t=20.3 \days$ (namely, 20.3 days after the explosion that mimics the equatorial ejecta). 
   
\begin{figure}[ht!]
	\centering
	\includegraphics[trim=5.2cm 8.2cm 4cm 8.2cm ,clip, scale=0.8]{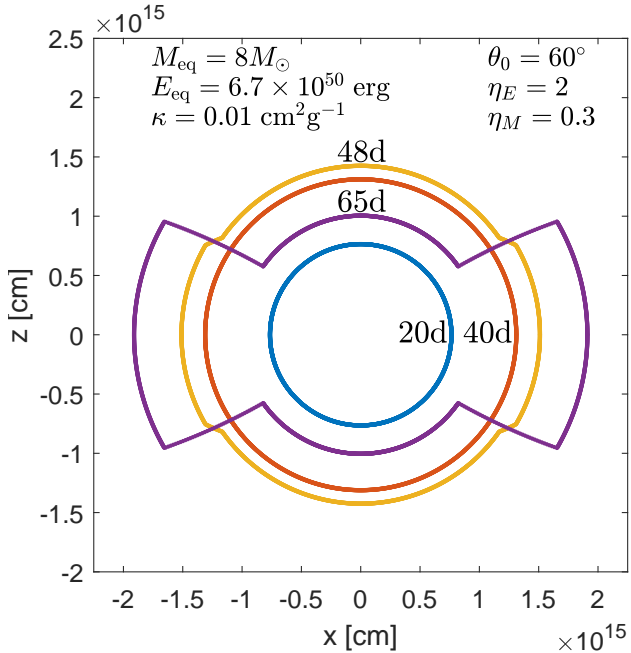}
	\caption{The shape of the photosphere in the meridional plane $y=0$ due to two late jets that interact with the SN ejecta, the late-fall case. The different colors represent different times as labeled in days. 
	At early times of $t \le 40 \days$ the photosphere is expanding as a spherically symmetric explosion. At $t_{\rm bipol} = 40 \days$ the two opposite polar bubbles, within an angle of $\theta_0=60^{\circ}$ from the symmetry axis, start to expand in the two polar directions. In the figure the expansion of the bubbles is not noticeable, because the polar and equatorial ejecta propagate at a similar rate. A short time later (a few days) the polar ejecta is already optically thin, and therefore the polar photosphere collapses at a very short time after the bubbles break out of the ejecta (seen here at 48 and 65 days).
	}
	\label{fig:shapes_late_fall}
\end{figure}

In Fig.~\ref{fig:shapes_late_fall} we present the shape of the photosphere in the meridional plane $y=0$ of our late-fall case at four times.
 
Let us elaborate on the role of the jets in this late-fall case. We assume that the jets displace mass from the polar regions to the sides. Our simple model cannot treat the displaced mass that forms a shell between the equatorial and polar outflows, and we ignore this mass in the present study. As well, the jets add energy to the mass that is left in the polar regions. This is the reason we assume that from $t_{\rm bipol} = 40 \days$ the polar regions contain less mass and more energy. Our simple assumptions require hydrodynamical simulations to reveal the more accurate flow structure. Since this polar faster expansion starts at a late time, for the parameters we use here within a short time the polar regions become optically thin and they do not contribute anymore to the luminosity that an equatorial observer measures. We also find that for a rapid fall in luminosity we need to take a low opacity. We take here $\kappa = 0.01 \cm^2 \g^{-1}$. Overall, the parameters we take for this case are demonstrative parameters that emphasize the effect we study here. Anyway, we expect such cases to be rare.  

\begin{figure}[ht!]
	\centering
	\includegraphics[trim=3.5cm 8cm 4cm 8cm ,clip, scale=0.6]{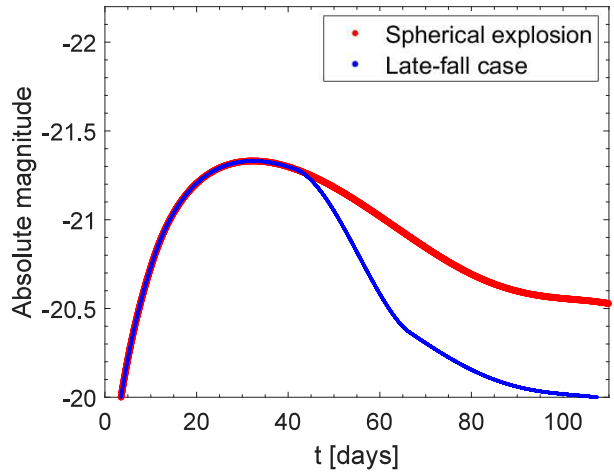}
	\caption{The light curve of the late-fall case (blue line) where the jets deform the spherical photosphere only from $t_{\rm bipol} = 40 \days$ (Fig.~\ref{fig:shapes_late_fall}). The thick-red line is the light curve of a spherical explosion with the properties of the equatorial ejecta as earlier cases. The parameters are $E_{\rm eq} = 0.667 \times 10^{51} \erg$, $M_{\rm eq} = 8 M_\odot$, $\eta_{\rm E}= 2 $, $\eta_{\rm M} = 0.3$, $\kappa = 0.01 \cm^2 \g^{-1}$ and $\theta_0=60^{\circ}$.
		}
	\label{fig:abs_magnitude_late_fall}
\end{figure}

Specifically, at $t=40 \days$ (the time we assume the jet-inflated bubbles break out from the photosphere of the spherical ejecta) we start to inflate the polar photosphere as we show in Fig.~\ref{fig:shapes_late_fall}. We use the same assumptions as in calculating the light curves that we present in Fig.~\ref{fig:abs_magnitude}, like a uniform photospheric temperature and a light curve of a spherical explosion that is given by the thick-red line in Fig.~\ref{fig:abs_magnitude}. Using equation (\ref{eq:lum}) we calculate the light curve that an observer in the equatorial plane would see. We present this light curve in Fig.~\ref{fig:abs_magnitude_late_fall}. We observe the rapid luminosity fall that starts at $t = 43 \days$ and ends at $t= 65 \days$, where we observe a break in the light curve that change from rapid fall to more moderate one.  
 
We emphasize that we chose the parameters to show that a late rapid luminosity fall is possible when late jets inflate polar bubbles. From the blue line in Fig.~\ref{fig:abs_magnitude_late_fall}, we learn that this is indeed the case. For many other parameters this will not occur. Our purpose is to point out that in many cases, but definitely not in the majority, jets might lead to geometrically-modified light curves. Here we concentrate on rapid luminosity falls. Other effects that jets might have on the light curves require separate studies (as in \citealt{KaplanSoker2020}).   

\section{An abrupt drop due to jet-inflated low-density bubbles}
\label{sec:Abrupt_drop}
\subsection{The low-density bubbles case}
\label{subsec:LowDensity}

The light curve of the hydrogen-poor luminous SN~2018don experiences a sharp decrease in its light curve after its peak \cite{Lunnan2019}. This motivates us to consider a somewhat different model of the bipolar ejecta than those we study in Sections \ref{sec:model} and \ref{sec:light_curve}.  We emphasize that at this stage we do not try to fit all properties of SN~2018don. We limit ourselves to show that geometrically-modified light curves might account for the abrupt drop in the light curve for an observer in the equatorial plane of the ejecta.   
 
We consider a flow structure where each of the two opposite jets inflates a bubble of hot low-density gas inside the SN ejecta. The radius of the photosphere at early times is as we show in Fig.~\ref{fig:shapes_early_times}. As the photospheric radius decreases in mass coordinate, it eventually reaches the low-density bubble interior that has a very low optical depth.
Therefore, the photosphere `collapses' to a very small radius, and an observer in the equatorial plane will stop getting radiation from the polar lobes. This causes a steep decrease in the cross section of the ejecta for this observer. 
     
We model this behavior as follows. For simplicity, we assume that the low density bubble is inside the break in the density profile (at $r = v_{\rm br} t$; equation \ref{eq:density_profile}). This implies that when the polar photospheric radius changes from being $r_1$ (equation \ref{eq:r1}) to $r_2$ (equation \ref{eq:r2}) we take the polar photosphere to be at a very small radius, practically smaller than the height of the equatorial ejecta $h_{\rm eq} = R_{\rm eq} \cos \theta_0$. 
The idealized geometrical structure of the jet-inflated bubbles causes a discontinues drop in the light curve. In reality it will be continues, but very sharp. 
         
\begin{figure}[ht!]
	\centering
	\includegraphics[trim=3.5cm 8cm 4cm 8cm ,clip, scale=0.6]{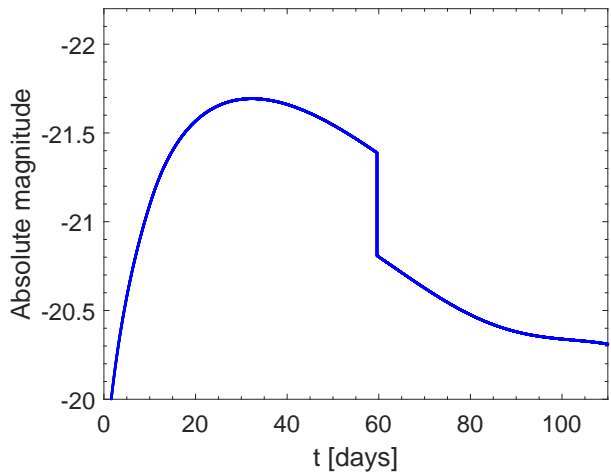}
	\caption{The light curve of the low-density bubbles case. In this case we assume that there are two low-density bubbles in the polar ejecta, one at each side of the equatorial plane. When the polar photosphere reaches the break in the density profile, the optical depth is very low, and the photosphere collapses abruptly to a very small radius. 
	This occurs for the parameters we use here at $t=59.6 \days$. We use equation (\ref{eq:lum}) to calculate the light curve, where $L_{\rm sp,eq}$ is the thick-red line in Fig.~\ref{fig:abs_magnitude} and \ref{fig:abs_magnitude_late_fall}, as in all previous cases.  
	The parameters we use here are $\theta_0=45^\circ$, $E_{\rm eq} = 1 \times 10^{51} \erg$, $M_{\rm eq} = 8 M_\odot$, $\eta_{\rm E}= 1.1 $, $\eta_{\rm M} = 0.4$ and $\kappa = 0.1 \cm^2 \g^{-1}$.
	}
	\label{fig:abs_magnitude_rapid_fall}
\end{figure}

For this case, the low-density bubbles case, we take the polar outflow to be within a half opening angle of $\theta_0=45^\circ$ from the symmetry axis (compared to $\theta_0=60^\circ$ in previous cases). For the other parameters we take $E_{\rm eq} = 1 \times 10^{51} \erg$, $M_{\rm eq} = 8 M_\odot$, $\eta_{\rm E}= 1.1 $, $\eta_{\rm M} = 0.4$ and $\kappa = 0.1 \cm^2 \g^{-1}$.
We take the light curve of a corresponding spherical explosion as before (the thick-red line in Fig.~\ref{fig:abs_magnitude}). 
We present the light curve of the low-density bubbles case in Fig.~\ref{fig:abs_magnitude_rapid_fall}.

\subsection{The light curve of SN~2018don}
\label{subsec:SN2018don}
\begin{figure}[ht!]
	\centering
	\includegraphics[trim=0.95cm 12cm 3.5cm 0.5cm ,clip, scale=0.55]{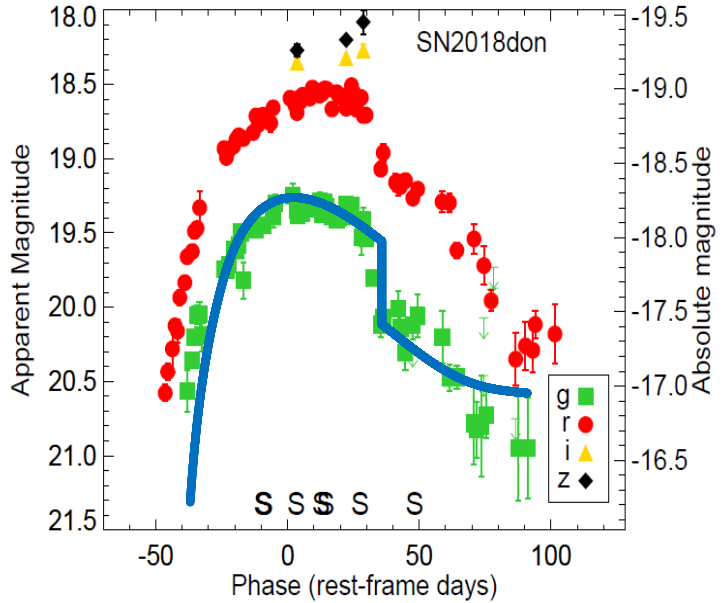}
	\caption{The red and green points are the light curves of SN~2018don in the r-band and g-band, respectively, as reported by \cite{Lunnan2019}. The blue line is the light curve of the low-density bubbles case as we present in Fig.~\ref{fig:abs_magnitude_rapid_fall}, shifted horizontally to match the time of the observed peak, and vertically by $+3.4$~mag. As we discuss in the text the actual vertical shift is only $+0.6$~mag. 
	}
	\label{fig:data_abs_mag_2008don}
\end{figure}

We present the light curve of SN~2018don as reported by \cite{Lunnan2019} in Fig.~\ref{fig:data_abs_mag_2008don}. 
The puzzle of this light curve is the abrupt drop by 0.7~mag, in both {\it g} and {\it r} bands. It occurs at about 35 days past maximum light. As well, SN~2018don is a superluminous CCSN (SLSN), that has other atypical properties, such as a red color and a massive host galaxy \cite{Lunnan2019}. 

To demonstrate that the low-density bubbles case might account for the abrupt drop in the light curve of SN~2018don, we start with the light curve that we present in Fig.~\ref{fig:abs_magnitude_rapid_fall} (where the model parameters are given). We move the light curve horizontally to match the time of the peaks of the g-band and of our light curve (this is a technical shift). 

We move our light curve vertically by $+3.4$~mag relative to that in Fig.~\ref{fig:abs_magnitude_rapid_fall}, to match the 
uncorrected g-band as given in Fig.~\ref{fig:data_abs_mag_2008don}. We note though, that \cite{Lunnan2019} estimate the extinction to SN~2018don, and after correcting the g-band, the maximum luminosity of SN~2018don is $-20.1$~mag. Namely, practically we need to move our graph vertically by only $+0.6$~mag, i.e., reduce the luminosity by a factor of 1.7. 

It is evident from Fig.~\ref {fig:data_abs_mag_2008don} that the abrupt drop in our low-density bubbles case pretty well reproduces the abrupt drop in the light curve of SN~2018don. 

Our simple model is not yet at a stage of comparing the light curve in different bands, nor to examine other viewing angles (see Section \ref{sec:summary}).

\section{Summary}
\label{sec:summary}

Under the assumption that jets explode many (or even all) CCSNe (section \ref{sec:intro}), we set the goal to examine one aspect of jet-inflated polar bubbles. In minority of cases the last jets that the NS (or BH) launches might be strong enough to inflate large polar bubbles, one on each side of the equatorial plane. The jets deposit energy in the polar regions, and therefore the polar bubbles expand faster that the equatorial ejecta. 
Specifically in this study, we examined the influence of such faster and lower density polar ejecta on the light curve that an observer in the equatorial plane would see. 
   
{{{{We built three simple axisymmetrical ejecta structures (models), all with faster polar outflows, i.e., bipolar structures. We calculated the light curves of four cases, as in the first model we calculated for two values of the explosion energy, }}}} the high-energy and low-energy cases (sections \ref{subsec:geometriccal} and \ref{subsec:lightcurve}). The other two models are the late fall case (section \ref{subsec:Shorter_fall}), and the low-density bubbles case (section \ref{sec:Abrupt_drop}). 

The basic process is as follows. At early times the photosphere in the polar ejecta is at larger radii. Because of the faster expansion and lower mass, the optical depth in the polar ejecta is lower, and at late times the photosphere recedes faster in the polar directions. Eventually, the polar ejecta is hidden behind the equatorial ejecta for an observer in the equatorial plane. We demonstrated this for the high energy case in Fig. \ref{fig:shapes_early_times}.

Under the assumption of a black body radiation with a uniform temperature (that changes with time) across the photosphere, we calculated the ratio of the luminosity of our geometrically modified light curves, to that of a spherical explosion with a radius equals to the radius of the equatorial ejecta in each case. This ratio is  the ratio of the cross section of the photosphere that an observer in the equatorial plane sees, to the cross section of a sphere with the radius of the equatorial ejecta (Fig. \ref{fig:area_ratio} for the high energy case and the low energy case). We took the light curve of SN~2007bi (from The Supernova Catalog; \citealt{SN_catalog}) to be that of a spherical explosion (thick-red line in Figs. \ref{fig:abs_magnitude} and \ref{fig:abs_magnitude_late_fall}). 
The desired geometrically light curve is given then by equation (\ref{eq:lum}). 

We present the four geometrically modified light curves of the four cases in Figs. \ref{fig:abs_magnitude}, \ref{fig:abs_magnitude_late_fall}, and \ref{fig:abs_magnitude_rapid_fall}, respectively.
In all cases the observer is in the equatorial plane. 

We point out that because of the simple models we use here, e.g., we have no radiative transfer calculations, we cannot examine yet the light curve for observers that are outside the equatorial plane. In particular, we will have to include radiative transfer from the equatorial ejecta to the polar directions, because we expect the equatorial ejecta to radiate substantially toward the polar directions after the collapse of the polar ejecta. For calculating the light curve as an observer outside the equatorial plane would see, we will have to construct more sophisticated models. 
{{{{ We can only estimate that the rapid collapse of the photosphere in the polar direction will expose relatively hot gas that radiate at shorter wavelengths compared to the rest of the supernova. This will lead to a `blue outburst'. Namely, a polar observer might see a late blue peak in the light curve that lasts for about several days to few weeks. }}}}

We summarize our main results as follows.  
Compared to the spherical SN light curve, 
The high energy and low energy cases light curves have  higher maximum luminosities than the light curve of the spherical ejecta. As well, they have a more rapid rise to the maximum and a steeper decline from maximum (Fig. \ref{fig:abs_magnitude}). The high energy case experiences an earlier luminosity rise and drop, making the decline in its light curve  more distinct.

The late fall case (Fig. \ref{fig:abs_magnitude_late_fall}) shows that a photosphere that suffers a late geometrically modification experiences a late rapid decline. 

In the low-density bubbles case, we where motivated by the puzzling light curve of SN~2018don that has sudden drop in its light curve (Fig. \ref{fig:data_abs_mag_2008don}). We showed that with the assumption of jet-inflated low-density bubbles (section \ref{subsec:LowDensity}), we could obtain an abrupt drop in the light curve. 
We could pretty well fit this light curve (blue line in Fig.  \ref{fig:data_abs_mag_2008don}) to the light curve of SN~2018don. 

In all the cases above we showed that when we consider a bipolar ejecta, i.e., those with faster polar ejecta,  
we find that an observer in the equatorial plane (of the ejecta) would measure a more rapid luminosity drop in the light curve, even an abrupt one, relative to a spherical ejecta. This property could serve to reveal the influence of strong jets in a minority of CCSNe. In particular, we suggest that the ejecta of SN~2018don was shaped by strong jets.

\section*{Acknowledgments}
{{{{ We thank Omer Bromberg, Ari Laor, and an anonymous referee for useful comments. }}}} This research was supported by a grant from the Israel Science Foundation and a grant from the Asher Space Research Fund at the Technion.

\label{lastpage}
\end{document}